\DocumentMetadata{} 
\documentclass[sigconf]{acmart} 
\AtBeginDocument{%
  \providecommand\BibTeX{{%
    \normalfont B\kern-0.5em{\scshape i\kern-0.25em b}\kern-0.8em\TeX}}}

\copyrightyear{2025}
\acmYear{2025}
\setcopyright{rightsretained}
\acmConference[CHI EA '25]{Extended Abstracts of the CHI Conference on Human Factors in Computing Systems}{April 26-May 1, 2025}{Yokohama, Japan}
\acmBooktitle{Extended Abstracts of the CHI Conference on Human Factors in Computing Systems (CHI EA '25), April 26-May 1, 2025, Yokohama, Japan}\acmDOI{10.1145/3706599.3720094}
\acmISBN{979-8-4007-1395-8/2025/04}

\usepackage{multirow}

\definecolor{cb_orange}{rgb}{1.0,0.51,0.0}
\definecolor{cb_blue}{rgb}{0.22,0.49,0.72}
\definecolor{cb_green}{rgb}{0.3,0.67,0.29}
\definecolor{cb_red}{rgb}{0.89,0.1,0.11}
\definecolor{cb_purple}{rgb}{0.6, 0.31, 0.64}
\definecolor{cb_brown}{rgb}{0.6, 0.4, 0.2}
\definecolor{cb_crimson}{rgb}{0.86, 0.08, 0.24}

\begin{document}

\title[Enhancing User Performance and Human Factors through VG in AR Assembly Tasks]{Enhancing User Performance and Human Factors through Visual Guidance in AR Assembly Tasks}



\author{Leon Pietschmann}
\orcid{0000-0001-6069-4567}
\authornote{Corresponding author: LHP30@cantab.ac.uk}
\affiliation{%
  \institution{University of Cambridge}
  \institution{Harvard University}
  \city{Cambridge}
  \country{UK/USA}
}

\author{Michel Schimpf}
\orcid{0009-0006-8864-2172}
\affiliation{%
  \institution{University of Cambridge}
  \city{Cambridge}
  \country{UK}
}

\author{Zhu-Tian Chen}
\orcid{0000-0002-2313-0612}
\affiliation{%
  \institution{Harvard University}
  \city{Cambridge}
  \country{USA}
}

\author{Hanspeter Pfister}
\orcid{0000-0002-3620-2582}
\affiliation{%
  \institution{Harvard University}
  \city{Cambridge}
  \country{USA}
}

\author{Thomas Bohné}
\orcid{0000-0001-5986-8638}
\affiliation{%
  \institution{University of Cambridge}
  \city{Cambridge}
  \country{UK}
}

\renewcommand{\shortauthors}{Pietschmann et al.}

\begin{abstract}
This study investigates the influence of Visual Guidance (VG) on user performance and human factors within Augmented Reality (AR) via a between-subjects experiment. VG is a crucial component in AR applications, serving as a bridge between digital information and real-world interactions. Unlike prior research, which often produced inconsistent outcomes, our study focuses on varying types of supportive visualisations rather than interaction methods. Our findings reveal a 31\% reduction in task completion time, offset by a significant rise in errors, highlighting a compelling trade-off between speed and accuracy. Furthermore, we assess the detrimental effects of occlusion as part of our experimental design. In addition to examining other variables such as cognitive load, motivation, and usability, we identify specific directions and offer actionable insights for future research. Overall, our results underscore the promise of VG for enhancing user performance in AR, while emphasizing the importance of further investigating the underlying human factors.
\end{abstract}

\ccsdesc[500]{Human-centered computing~Empirical studies in visualization}
\ccsdesc[500]{Human-centered computing~Mixed / augmented reality}
\ccsdesc[500]{Human-centered computing~User studies}

\keywords{Visual Guidance, AR/VR/XR/Immersive, Instructions, User Study, Experiment}

\begin{teaserfigure}
  \centering
  \includegraphics[width=0.99\textwidth]{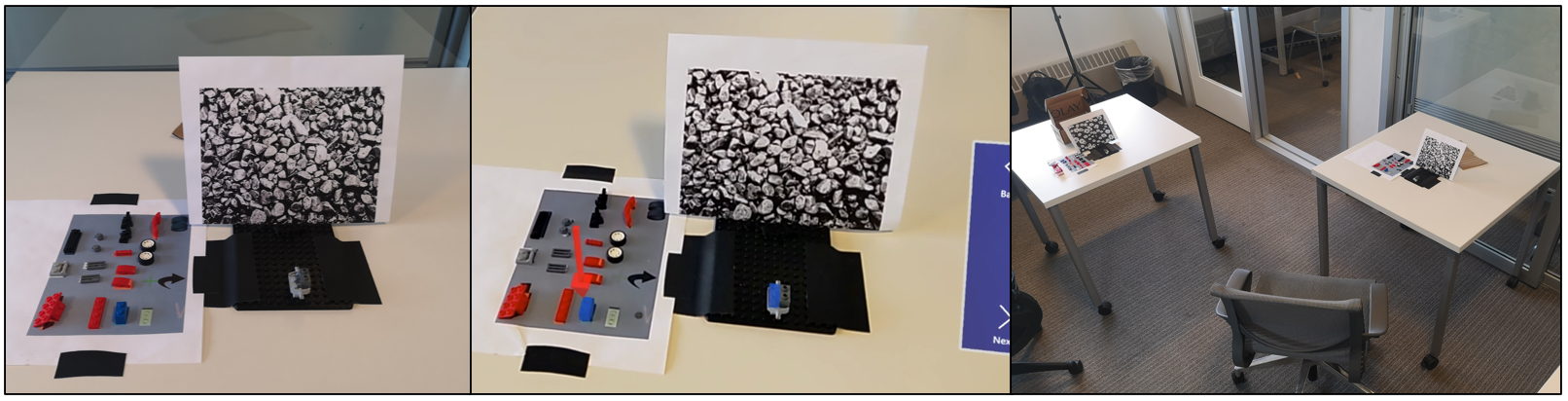}
  \caption{From left to right: raw assembly setup, overlaid Visual Guidance (arrow and holograms), two of the three workstations}
  \Description{This figure consists of three sub-images showing an experimental setup on a table. The setup includes LEGO bricks placed on a paper sheet with printed guidelines and a black-and-white printed image of gravel in the background. The leftmost and center images provide close-ups of the experimental station, while the rightmost image shows the wider room setup with two identical experiment stations in a controlled office setting.}
  \label{fig:teaser}
\end{teaserfigure}


\maketitle

\section{Introduction}
Technologies within the Extended Reality (XR) spectrum, encompassing Augmented and Virtual Reality (AR/VR), 
offer users an integrative experience that merges real and virtual components. 
Such innovations have the capacity to profoundly influence sectors such as manufacturing~\cite{Jasche2021,Bttner2020} or education~\cite{ARchemist,DementiaEyes}. 
Through the integration of visual aids directly into the user's visual field, 
XR tools can be utilised to enhance efficiency during training and actual operations, for instance, by diminishing task completion time or minimising errors in execution~\cite{yang_2019,blattgerste_2018,Pietschmann2022ICHMS,Pietschmann2023VIS,Pietschmann2023THW}.
Extended Reality Visual Guidance (XRVG), discussed below, holds the potential to further improve the efficiency of implementing XR technologies, particularly for industrial environments~\cite{Pietschmann2022ICHMS,Pietschmann2023VIS,pietschmann2024PhDThesis}.

While previous studies mostly focus on comparing different media and XR types~\cite{Jasche2021}, we found notably fewer comparing the impact of different visualisation types on user performance in XR. 
Still, the empirical evidence on the impact of XRVG remains mixed.
Moreover, some publications described that participants were criticising the occlusion of real-world objects by virtual elements in XR visualisations~\cite{Jasche2021,Muoz2019}. 
As far as we know, the effects of occlusion in an XRVG context were not measured in any study before.

This study aims to quantify the impact of visualisation type and occlusion in XR on user performance and human factors.
Specifically, we conducted a between-subject experiment evaluating different Augmented Reality visualisations and an Occlusion Avoidance Feature (\textbf{OAF}) on an assembly task. 
In the experiment, the participants were asked to assemble a Lego car. Depending on the treatment group, instructions were shown on a virtual slide or using XRVG.
In particular, we examined the task completion time/total time to completion (\textbf{TTC}), error rate, cognitive load, usability, motivation, and perceived helpfulness.
With this research, we hope to contribute to an improved understanding of people in relation to XR and, more generally, to the field of XR visualisation by
(1) showing how XRVG can decrease TTC while increasing error rate and usability revealing an interesting conflict between TTC and mistakes made,
(3) by demonstrating that there is no effect on cognitive load, motivation and perceived helpfulness,
(2) and quantifying the effect of occlusion in XRVG by testing the OAF, which results in a decreased TTC.


\section{Background}
Extended Reality Visual Guidance (XRVG) refers to embedding real-time situated visualisations within the user's field (for example, the red arrow in the second subfigure in \autoref{fig:teaser}) by using Augmented or Virtual Reality head-mounted displays (AR/VR HMDs)~\cite{Pietschmann2022ICHMS,Pietschmann2023VIS}. 
Few studies so far have explored the impact of comparing different forms of Visual Guidance, especially in an industrial assembly context~\cite{Satkowski2021,Burova2020,Pietschmann2023VIS,Williams2021}. 
The overall consensus on the potential of XRVG for supporting users complete their tasks -- especially procedural knowledge-intensive sequential tasks --  is positive~\cite{Pietschmann2023VIS,Pietschmann2022ICHMS}. 


However, the associated literature has reported mixed empirical evidence. While \cite{Seeliger2023} reported a reduction in TTC, \cite{Jasche2021} reported finding no significant impact. While \cite{Seeliger2023,Jasche2021} reported a reduction in error rates, \cite{Smith2020} reported finding no significant impact. For the underlying human factors, previous studies found an anticipated reduction~\cite{lampen_2019,yang_2019,Seeliger2023,Smith2020}, an increase~\cite{baumeister_2017,Blattgerste2017}, and no significant differences~\cite{blattgerste_2018,Yang2020,Jasche2021} in cognitive load. Similarly, previous studies found a positive~\cite{Seeliger2023,Smith2020} and no significant effect on usability~\cite{Jasche2021} and motivation~\cite{Yang2020,Farr2023}. These conflicting results may be associated with the lack of a systematic approach to investigating XR Visual Guidance, the underexplored impact of occlusion, and potential novelty effects, stemming from participants' unfamiliarity with XR hardware, thus causing opposing influences on the results~\cite{Pietschmann2023VIS,makransky_2017}.

Despite receiving increased attention, systematic approaches to XRVG, such as the framework proposed by \cite{Pietschmann2023VIS}, and their implementation as part of a between-subject design remain scarce. Moreover, existing studies are yet to use AR technology, thus adding a physical interaction layer to the HCI process. Lastly, we were not able to identify any studies exploring and/or quantifying the impact of occlusion in XRVG contexts as part of a between-subject experiment.

In this study, we explore the impact of XRVG on user performance and human factors in an industrial AR assembly context while employing similar between-subject experimental design also supported by the XRVG framework used in \cite{Pietschmann2023VIS} and simultaneously investigating the impact of occlusion. Concretely, we investigate six hypotheses, namely detecting a difference in TTC, number of mistakes, cognitive load, usability, motivation, and perceived helpfulness of the VG between the groups (\textbf{Hypotheses H1-H6} respectively).


\begin{itemize}
    \item \textbf{H1}: TTC significantly differs between the different visualisation types.
    \item \textbf{H2}: The number of mistakes significantly differs between the different visualisation types.
    \item \textbf{H3}: Cognitive load significantly differs between the different visualisation types.
    \item \textbf{H4}: Usability significantly differs between the different visualisation types.
    \item \textbf{H5}: Motivation significantly differs between the different visualisation types.
    \item \textbf{H6}: Perceived helpfulness of the Visual Guidance significantly differs between the different visualisation types.
\end{itemize}

\section{Methodology}

A mixed-method, between-subject study design was chosen to pursue the aforementioned objectives for this study. 
Besides investigating the effect of the implemented Visual Guidance on user performance and human factors, 
we also investigated the applicability of the XRVG framework proposed by~\cite{Pietschmann2022ICHMS} in an AR context as well as explore how the previously mentioned challenge around occlusion of visual cues could be overcome. 

\subsection{Occlusion Avoidance Feature (OAF)}
\label{OAF}

To mitigate the occluding effect of the Visual Guidance elements, we developed and tested an Occlusion Avoidance Feature (OAF). Previous studies pointed out that visual cues may obstruct the users' field of view during the picking and placement process~\cite{Pietschmann2023VIS}, necessitating methods to reduce occlusion. The OAF was designed to deactivate the supporting VG when participants began assembling parts, preventing obstructions during this process.
It works by constantly monitoring the positions of virtual objects and participants' hands, activating or deactivating objects based on a 10cm distance threshold to avoid accidental changes. 

The OAF aims to support the cognitive process leading up to the execution of the motoric process (i.e. deciding which part to pick or where to place). Once the user makes a decision and begins the picking and placing process, the OAF disables the VG to avoid occlusion, thus enhancing the execution of the physical task. While XRVG supports the cognitive process, the OAF focuses on improving the motoric process, reducing the negative impact of VG elements, and ultimately enhancing overall user performance and human factors.


\subsection{Study Design}

To examine, quantify, and compare the impact of XRVG on user performance and human factors, and explore its transferability, the study design included three groups: a control group (\textbf{G1}) without VG, a treatment group (\textbf{G2}) with full VG following the XRVG framework ~\cite{Pietschmann2022ICHMS,Pietschmann2023VIS} (i.e. Gaze Guidance [\textbf{GG}], Object Identification [\textbf{OI}], and Action Guidance [\textbf{AG}]), as well as a third group (\textbf{G3}), which incorporated a newly developed Occlusion Avoidance Feature (\textbf{OAF}) alongside G2 to address occlusion effects.

\begin{figure}[htb]
    \centering
    \includegraphics[width=0.47\textwidth]{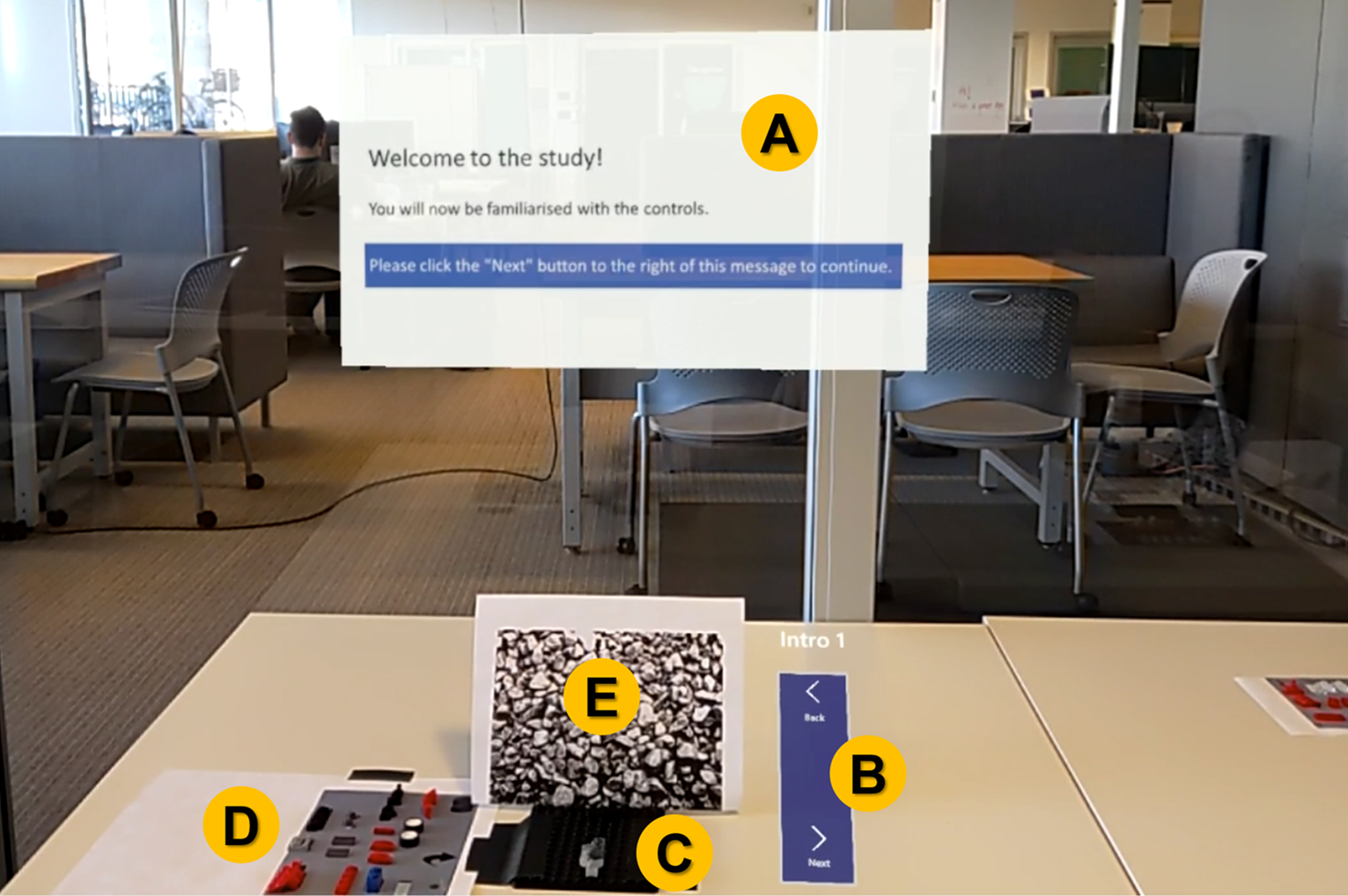}
    \caption[Experimental setup for the AR study]{Experiment setup and implementation of AR overlays, specifically instruction screen (A), navigation menu (B), main workpiece (C), parts area (D), and tracking target (E)}
    \label{fig:Study3Environment}
    \Description{This image showcases the AR interface as seen through the headset. The instructional message (A) guides the participant, while the "Next" button (B) allows for navigation. The LEGO baseplate (C) serves as the placement area for pieces, and the printed parts area (D) includes an overlay of LEGO components with the physical pieces placed on top. The printed tracking target (E), a black-and-white gravel background, is used for spatial alignment in the AR system.}
\end{figure}

A between-subjects design was used to avoid learning effects, with all participants wearing AR HMDs to ensure consistency. 
All groups were given the same task and the same instructions. 
For all groups, a virtual monitor displayed assembly instructions, resembling Lego’s official manual, and a navigation menu allowed participants to move between steps. The AR content was locked into place by tracking the main target, ensuring correct display. For all groups, a large virtual monitor above Workstation 1 displayed the (assembly) instructions based on Lego’s official manual. \autoref{fig:Study3Environment} (A) shows the concrete implementation within the AR environment. 
Furthermore, participants had a navigation menu to progress or return to the next or previous steps if necessary (B) which they selected with their finger. The virtual screen as well as the navigation menu hovered in spatial proximity to the assembly area and the main workpiece (C), as well as the parts area (D). All AR content was locked into place by tracking the main target (E), which helped the AR device to identify the assembly area and correctly display the AR content.

\begin{figure}[htb]
    \centering
    \includegraphics[width=0.47\textwidth]{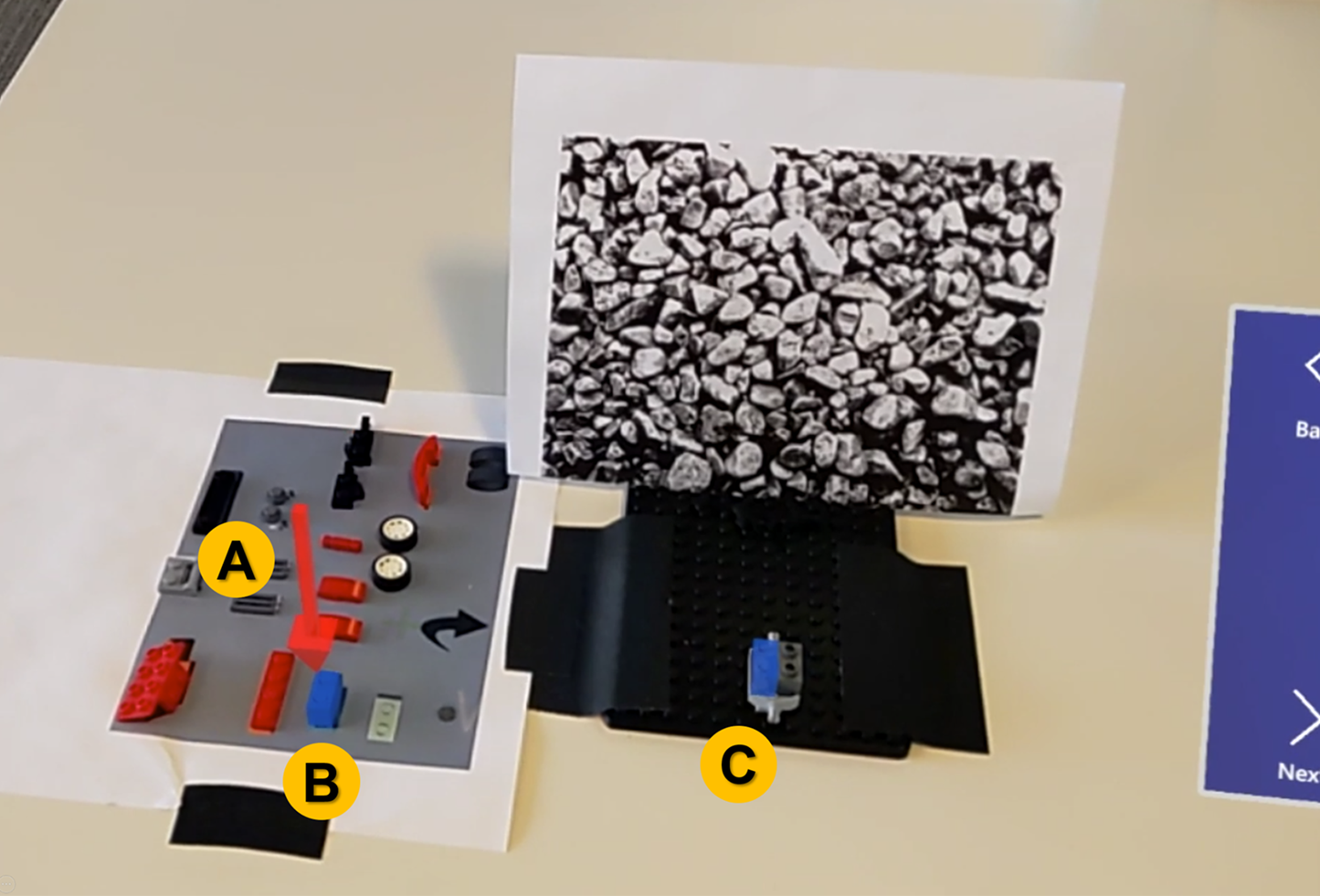}
    \caption[Implementation of XRVG for AR study]{Implementation of Object Identification (Arrow A and Hologram B) and Action Guidance (Hologram C)}
    \label{fig:Study3XRVG}
    \Description{A close-up of the workspace highlights labeled LEGO pieces and their designated placement areas. This image provides a detailed view of how the experimental task is structured, showing both the printed reference sheet and physical LEGO components.}
\end{figure} 

While G1 only received instruction screens and navigation menu (A in \autoref{fig:Study3Environment}), G2 and G3 featured additional XRVG components: following \cite{Pietschmann2023VIS}, GG was implemented with a large red arrow to guide participants to the next workstation, OI highlighted the next part to pick with a red arrow (A in \autoref{fig:Study3XRVG}) and hologram overlay (B), and AG showed the target location and orientation of the part (C). The AR application was developed using Unity, Vuforia, and Microsoft’s Mixed Reality Toolkit and deployed using a Microsoft HoloLens 2.

The application was built locally, sideloaded onto the device and installed including all dependent packages prior to the start of the study. The benefit of this method was that, once complete, the application was independent from outside computers, always followed the same procedure, and could be started by simply pressing a button, which added to the speed, reliability, and resilience of the study procedure. The device was restarted frequently to clear the local cache, minimise tracking errors, and ensure overall device performance.

\subsection{Participant Recruiting and Selection}

An ex-ante power analysis supported by \textit{G*Power} 3.1~\cite{Faul2009} indicated a sample size of over 27. While this number seemed low for a between-subjects approach with three groups, the large effect sizes reported in a comparable study by \cite{Pietschmann2023VIS} supported this power analysis. Randomization allowed for efficient participant allocation, and eligibility criteria included being over 18, with no major vision impairments. 

\begin{figure}[htb]
    \centering
    \includegraphics[width=0.47\textwidth]{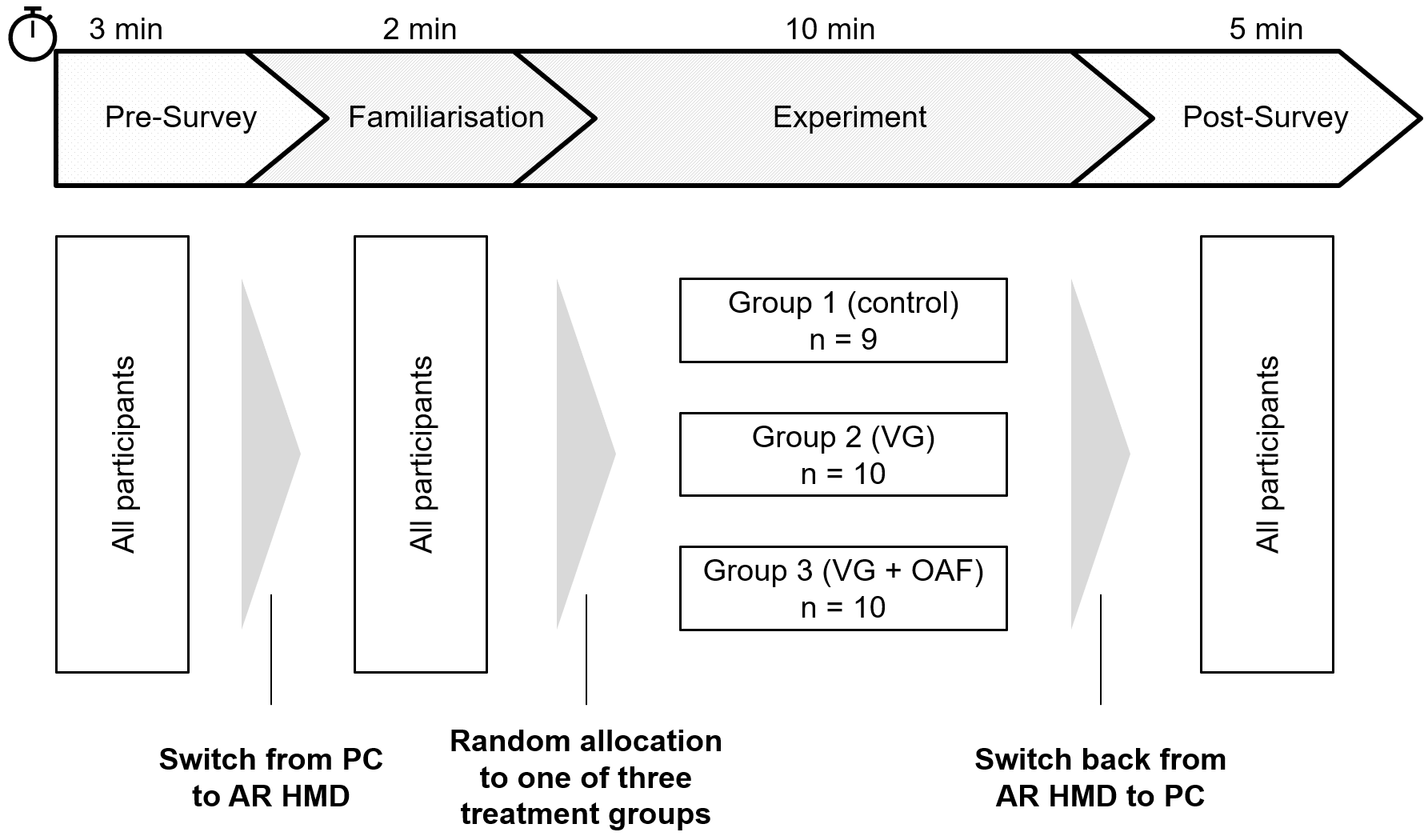}
    \caption{Experimental procedure, including treatment groups, and participants per treatment group}
    \label{fig:Study3Process}
    \Description{This figure illustrates the experimental procedure using a timeline and flowchart. Participants begin with a Pre-Survey (3 minutes), followed by Familiarisation (2 minutes) using an AR headset. They are then randomly allocated into one of three groups: Group 1 (Control, n=9), Group 2 (VG - Visual Guidance, n=10), or Group 3 (VG + OAF - Visual Guidance with On-screen Augmented Feedback, n=10). The Experiment phase lasts 10 minutes, and the study concludes with a Post-Survey (5 minutes) after switching back from the AR headset to a standard PC.}
\end{figure}

\textbf{Procedure:} The study procedure involved a pre-survey, a familiarization phase, a main assembly task, and a post-survey (illustrated in \autoref{fig:Study3Process}), typically completed within 30 minutes. Participants were briefed on the experiment, familiarized with the AR HMD, and instructed to assemble a Lego car (set 31055) as fast as possible in 63 steps across three workstations. At the start of the experiment, participants were familiarised with the controls, interaction methods, and specifics about the virtual environment before they started the timed main assembly. This pre-training principle was incorporated to mitigate potential novelty effects on the results~\cite{Pietschmann2023VIS, meyer_2019}. After completing the task, participants progressed to the post-survey, received their incentive, and were discharged.

\textbf{Measurements:} Measurements included informed consent, demographics, time to completion, mistakes, cognitive load, usability, motivation, perceived helpfulness of visual support, and qualitative feedback. The latter was collected at various points during pre- and post-surveys to allow participants to elaborate on their answers and to increase context sensitivity. Data were collected via Qualtrics and locally analysed using the SPARC software \cite{sparc2024}.


\newcommand{\TTCeta}{0.481}
\newcommand{\TTCCohensf}{0.926}
\newcommand{\Mistakeseta}{0.446}
\newcommand{\MistakesCohensf}{0.805}
\newcommand{\CLeta}{0.031}
\newcommand{\CLCohensf}{0.032}
\newcommand{\SUSeta}{0.236}
\newcommand{\SUSCohensf}{0.308}
\newcommand{\IMIeta}{0.043}
\newcommand{\IMICohensf}{0.044}
\newcommand{\VisQeta}{0.065}
\newcommand{\VisQCohensf}{0.069}

\section{Results}
\label{sec:results}

\begin{figure}[htb]
    \centering
    \includegraphics[width=0.475\textwidth]{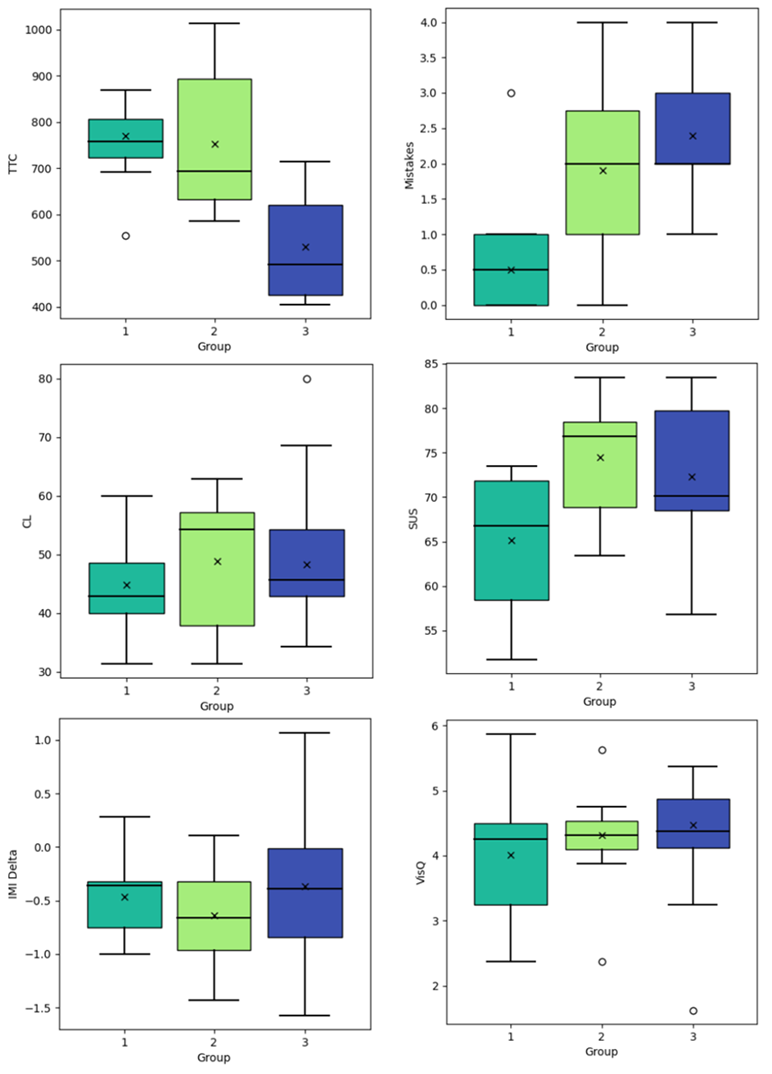}
    \caption{Boxplots for TTC, mistakes, CL, usability, motivation, and perceived helpfulness per group}
    \label{fig:Study3Boxplots}
    \Description{This figure consists of six boxplots comparing different metrics across the three experimental groups. The Time to Completion (TTC) shows that Group 3 had the lowest median completion time. The Mistakes metric indicates that the control group made the fewest errors, while Groups 2 and 3 exhibited higher mistake rates. Cognitive Load (CL) was slightly higher for Groups 2 and 3 compared to the control group. The System Usability Score (SUS) was higher for Groups 2 and 3, suggesting better perceived usability of the AR interface. The IMI Delta (Intrinsic Motivation Increase/Change) was similar for Groups 2 and 3. Finally, the Visual Quality Score (VisQ) was rated higher for Groups 2 and 3 than for the control group, indicating that participants found the AR interface visually effective.}
\end{figure}

A total of 29 participants were included in this study. Based on the analysis of the dependent variables (i.e. TTC, mistakes, CL, usability, motivation, and helpfulness of VG), all distributions were tested for normality and homoscedasticity. The assumption of normality was invalidated only once for the distribution of mistakes made, which was analysed using the Kruskal-Wallis (KW) test~\cite{Kruskal1952}. Furthermore, the distribution of VisQ failed the Levene test and was consequently analysed using the Welch-ANOVA. All other distributions were analysed using ANOVA, as they passed all tests for both the assumption of normality and homoscedasticity. Due to the high amount of normal distributions, it was decided to compare and calculate the differences between the groups using the groups’ means $\mu$~\cite{Brown1974,Kruskal1952}. For an overview of the results see \autoref{tab:Study3Normality} and \autoref{fig:Study3Boxplots}.

\begin{table}[htb]%
\centering%
\scriptsize 
\caption{Test results for normality and homoscedasticity for each of the six dependent variables}%
\begin{tabular}{l|ll|ll|ll}%
    \hline%
    \multirow{2}{*}{Dep. Variables}&\multicolumn{2}{c}{Shapiro{-}Wilk}&\multicolumn{2}{c}{Levene}&\multicolumn{2}{c}{Analysis Method}\\%
    &Min p&Result&Min p&Result&Analysis&Post hoc\\%
    \hline%
TTC&0.116&Pass&0.096&Pass&ANOVA&Tukey\\%
Mistakes&<0.001&Fail&0.147&Pass&Kruskal-Wallis&Dunn\\%
CL&0.09&Pass&0.49&Pass&ANOVA&Tukey\\%
SUS&0.351&Pass&0.712&Pass&ANOVA&Tukey\\%
IMI Delta&0.652&Pass&0.225&Pass&ANOVA&Tukey\\%
VisQ&0.811&Pass&0.048&Fail&Welch-ANOVA&Games-Howell\\%
\hline%
\end{tabular}%
\label{tab:Study3Normality}
\end{table}

\textbf{Time to Completion:} H1 states that the time to completion differs significantly between the different visualisation types. H1 was confirmed since the results and the ANOVA test ($F=11.576$, $p<0.001$, $\eta^2=\TTCeta$, $f^2=\TTCCohensf$) demonstrate that TTC differs significantly between the visualisation types. Moreover, the Tukey post hoc test shows a significant difference between G1 and G3 ($p<0.001$) as well as G2 and G3 ($p=0.001$).

\textbf{Number of Mistakes:} \label{MistakesResultsChap6} H2 states that the average number of mistakes differs significantly between the visualisation types. H2 was confirmed since the results and the non-parametric KW test ($H=12.042$, $p=0.002$, $\eta^2=\Mistakeseta$, $f^2=\MistakesCohensf$) demonstrate that the number of mistakes differs significantly between the visualisation types. Moreover, the Dunn post hoc test shows a significant difference between G1 and G2 ($p=0.023$) as well as G1 and G3 ($p=0.002$). 

\textbf{Cognitive Load:} H3 states that cognitive load differs significantly between the different visualisation types. H3 was not confirmed since the results and the ANOVA test ($F=0.402$, $p=0.673$, $\eta^2=\CLeta$, $f^2=\CLCohensf$) demonstrate that cognitive load does not differ significantly between the visualisation types. Applying the Tukey post hoc test did not yield any significant results either.

\textbf{Usability:} H4 states that usability differs significantly between the different visualisation types. H4 was confirmed since the results and the ANOVA test ($F=4.01$, $p=0.03$, $\eta^2=\SUSeta$, $f^2=\SUSCohensf$) demonstrate that usability differs significantly between the visualisation types. Moreover, the Tukey post hoc test shows a significant difference between G1 and G2 ($p=0.03$).

\textbf{Motivation:} H5 states that motivation differs significantly between the different visualisation types. H5 was not confirmed since the results and the ANOVA test ($F=0.577$, $p=0.568$, $\eta^2=\IMIeta$, $f^2=\IMICohensf$) demonstrate that motivation does not differ significantly between the visualisation types. Applying the Tukey post hoc test did not yield any significant results either.

\textbf{Perceived Helpfulness of Visual Guidance:} H6 states that the perceived helpfulness of VG differs significantly between the different visualisation types. H6 was not confirmed since the results and the Welch-ANOVA test ($F=0.552$, $p=0.588$, $\eta^2=\VisQeta$, $f^2=\VisQCohensf$) demonstrate that VisQ does not differ significantly between the visualisation types. Applying the Games-Howell post hoc test did not yield any significant results either. 

\textbf{Qualitative Feedback:} Most participants providing free-text answers were satisfied with their experience and the implemented XR Visual Guidance. Interestingly, several control group participants pointed out a need for Visual Guidance, unaware it was already implemented in other treatment groups.
Similarly, multiple participants from G2 expressed dissatisfaction with occlusion caused by the Visual Guidance and recommended a need for the Occlusion Avoidance Feature, again unaware this was included in G3.Regarding Gaze Guidance, several control group participants reported difficulty orienting themselves during the experiment, such as knowing what to do after completing the first workstation. This issue was absent in G2 and G3, which both featured a GG arrow pointing to the next workstation. However, one G3 participant suggested the GG feature be active at all times, not just after completing a workstation.

Overall, written feedback supported the XR Visual Guidance framework and highlighted the importance of the Occlusion Avoidance Feature. Complaints centred on tracking errors, mismatches between visuals and real components, and issues with the navigation menu. No participants reported problems with cybersickness, vision, or other health-related factors impeding their performance. Overall, the experiment and especially the XR Visual Guidance and the newly developed OAF were well-received.

\section{Discussion}

With H1, H2, and H4 confirmed, the empirical data reveal a significant effect of XR Visual Guidance on user performance in terms of time to completion and mistakes made, as well as usability. \autoref{tab:Study3Results} summarises the results for each hypothesis/dependent variable. 
Interestingly, most distributions satisfied the conditions for normality and homoscedasticity: only mistakes made exhibited a non-normal distribution, while VisQ was the only non-homoscedastic distribution. The results affirmed a significant difference between the individual visualisations for both TTC and the number of mistakes made. The overall differences between the groups for TTC and mistakes are strongly significant ($p<0.001$ and $p=0.002$ respectively). While the combination of Visual Guidance and the newly developed OAF was able to reduce TTC by as much as 240 seconds or 31\%, mistakes made surprisingly rose from an average of 0.5 (G1) to 2.4 (G3), which is an increase of 380\% compared to the control group.

\begin{table}[htb]
    \centering
    \scriptsize
    \caption{Overview of hypotheses tests, group differences (compared to control group), and effect sizes}
    \begin{tabular}{l|c c c}
    \toprule
    Hypothesis & Hypothesis confirmed? & Max Delta & Effect size \\
    (and dependent variable) & (p-value) & (vs G1) & ($\eta^2$) \\
    \midrule
    H1: TTC & Yes ($p<0.001$) & 31\% & \TTCeta~(large) \\
    H2: Mistakes & Yes ($p=0.002$) & (380\%) & \Mistakeseta~(large) \\
    H3: CL & No ($p=0.673$) & - & - \\
    H4: Usability & Yes ($p=0.030$) & 14\% & \SUSeta~(large) \\
    H5: Motivation & No ($p=0.568$) & - & - \\
    H6: VisQ & No ($p=0.588$) & - & - \\
    \bottomrule
    \end{tabular}
    \label{tab:Study3Results}
\end{table}


 \subsection{Impact of Visual Guidance} 

Interestingly, implementing Visual Guidance alone only slightly reduced the task completion time compared to the control group (2\% reduction between G1 and G2), whereas the additional implementation of the OAF accounted for the majority of the overall improvement (31\% reduction between G1 and G3). Furthermore, the standard deviation for TTC in G2 was notably larger than in the other groups. In combination with the qualitative feedback it appears that although the implemented Visual Guidance was initially perceived as helpful, it then proceeded to occlude at least some participants’ field of vision during the physical placement process. This detrimental effect seems to have been mitigated by implementation of the newly developed OAF, as indicated by both the analysis of the results and the participants’ qualitative feedback. In terms of mistakes made, the steep increase in mistakes seems counterintuitive at first. Additionally, the distribution of mistakes made was the only non-gaussian distribution, and the standard deviation for G2 and G3 was notably higher compared to the control group. Based on the qualitative feedback, three potential factors contributing to this were identified: 

\textbf{Occlusion:} Multiple participants reported occlusion to be a source of confusion and errors for them. It seems like some participants were particularly affected by the Visual Guidance obstructing the placement process, mainly when trying to locate the correct target position to place their part. This appears especially true for G2, where both the mean and median number of mistakes as well as standard deviation for both TTC and mistakes made was highest.

\textbf{Misalignments:} Slight misalignments in the perception of the visual overlays and the physical parts might have contributed to incorrect placements, as was reported by some participants. Previous studies have reported estimated tracking error for the HoloLens AR HMD of around 1cm~\cite{Yan2021}. Although the tracking error in this study was likely smaller—around 0.5 cm—even this margin is significant for small Lego parts, which can still fit together when slightly misaligned. The sharp increase in mistakes in G2 and G3 compared to G1 supports this hypothesis, as G1 lacked Visual Guidance components. However, potential misalignments were likely minor, as neither participants nor the research team who regularly calibrated and checked the setup, identified noticeable issues. Additionally, the mistakes did not follow a systematic pattern, making alignment problems with specific parts of the setup unlikely.

\textbf{Depth of Processing:} The main working hypothesis draws on the 'depth of processing' theory~\cite{craik_1972}, which suggests that overly simplifying a task by guiding users too effectively reduces cognitive load to the point where active or deep processing of instructions is hindered~\cite{yuvilergavish_2011}.  Visual Guidance aims to support the placement process by providing situated visualisations, minimizing the transfer of information from manual instructions to the physical task. This reduces the need for mental rotation, mitigates the split-attention effect, and lowers cognitive load.
However, reducing the required depth of processing for part placement appears to bypass the checks and controls needed for accurate execution. Whereas participants in G1 had to mentally interpret and verify instructions to ensure proper placement, those in G2 and G3 increasingly relied on Visual Guidance alone, eventually ceasing to refer back to the instruction screen. Some participants checked the instructions when unsure about part placement, but others misplaced parts without verifying, leading to mistakes.

The increase in mistakes may also stem from the cognitive overload hypothesis, suggesting that the placement process was overly complex, making it difficult for participants to identify the correct target location. As CL increased from G1 to G2 and G3, so did the number of mistakes. However, the association between CL and errors is weak—G2 had the highest CL but not the most mistakes—and the CL for G2 and G3 fell into the "somewhat high" range, which typically does not indicate cognitive overload. While this theory cannot be dismissed entirely, it seems unlikely here.

These findings on TTC are consistent with previous studies~\cite{lampen_2019,yang_2019,Seeliger2023,blattgerste_2018,Smith2020}, especially in terms of XRVG's beneficial effects on the picking and placement process. However, they contrast with studies \cite{baumeister_2017,Jasche2021,Yang2020} reporting negative or negligible impacts on TTC. This contrasting evidence may arise from ceiling effects in studies with low-complexity assembly designs \cite{baumeister_2017,Jasche2021}, where tasks were too simple to reveal XRVG’s benefits. As \cite{yang_2019} points out, low-complexity tasks allow participants to perform perfectly without support, potentially masking XRVG’s advantages or even hindering performance. Observations of the detrimental effects of occlusion-related challenges in this study further support this explanation and highlight the need to consider task complexity when evaluating XRVG.

The significant rise in mistakes is surprising and inconsistent with prior studies \cite{yang_2019,Jasche2021,tang_2003,Seeliger2023,Yang2020}, none of which reported such a sharp increase. One possible explanation is that the study design amplified the detrimental effects of the depth-of-processing trade-off on placement accuracy. This aligns with findings from \cite{Blattgerste2017}, where participants using XRVG with a HoloLens HMD showed a notable increase in placement errors (effect size $\eta^2=0.24$, which can be considered 'large') while attempting to locate correct assembly locations. This suggests that the interplay between depth of processing and XRVG implementation requires further investigation.


\textbf{Cognitive load:} H3 was not confirmed, as NASA-TLX scores revealed no significant differences in cognitive load between individual visualisations. Interestingly, the control group (G1) exhibited the lowest CL, despite having the highest TTC and the lowest average number of mistakes. In contrast, G2 and G3 had similar CL levels, falling within the upper range of the "somewhat high" category (scores between 44 and 49), but displayed significant differences in performance outcomes. This performance deficit for TTC could be due to cognitive underload, while for mistakes made a cognitive overload seems plausible. However, these interpretations are contradictory, as cognitive underload and overload cannot coexist simultaneously. Furthermore, the observed range of CL scores between 44 and 49 suggests no extreme cognitive load or ceiling/floor effects, and the similarity in CL across groups fails to account for the stark differences in performance. Additionally, the results did not cross the significance threshold ($p=0.67$).

The inconclusive findings on CL align with the mixed results reported in prior studies. While some \cite{lampen_2019,Seeliger2023,Smith2020,yang_2019} indicated a reduction in CL with XRVG, other studies \cite{baumeister_2017,Blattgerste2017} report its increase. This variability may stem from differences in experimental setups, as studies with designs closer to this experiment ~\cite{Jasche2021,Yang2020,blattgerste_2018} also report no effect on CL. These findings underscore the complexity of evaluating CL and its relationship with performance in XRVG-based tasks.

\textbf{Usability:} H4 was confirmed, as the SUS score differences for individual visualisations were significant. Groups G2 and G3 surpassed the critical usability threshold of 70 ~\citep[p. 120]{bangor2009determining}, indicating no major usability issues, while the control group (G1) averaged a lower SUS score of 65, suggesting poor usability. Usability for G2 and G3 was rated as good, whereas G1's usability was rated as low. However, attempts to correlate usability with performance revealed no consistent patterns. Although usability scores for G2 and G3 were higher than G1, TTC was similar for G1 and G2, with G3 achieving the lowest TTC. Likewise, the distribution of mistakes did not align with usability ratings. These findings mirror CL results, where significant performance differences lacked a clear relationship with usability scores.

Interestingly, the implementation of XRVG alone led to a significant increase in usability, despite all interaction methods being identical across groups. This aligns with prior studies ~\cite{Smith2020,Seeliger2023} that highlighted the positive impact of XR Visual Guidance on usability. Moreover, it underscores the importance of addressing usability challenges in XR systems, as previously identified by ~\cite{akayr_2017,Ibez2018,Radu2014,thees_2020}.

\textbf{Motivation:} H5 was not confirmed, as there was no significant difference in the IMI scores between the visualisations. Better performance did not appear to correlate with higher motivation, and motivation differences did not explain performance variations. Moreover, all IMI scores were negative, suggesting a general decrease in motivation from pre- to post-experiment. This result contradicts previous studies that reported increased motivation with XR technology use ~\cite{Freina2015,Makransky2019,Mouatt2020} and improved features ~\cite{Huang2020}. One possible explanation is that participants found the experience disappointing, though this contradicts much of the positive qualitative feedback and high usability scores.

However, the lack of significance and mixed descriptive statistics align with literature reporting inconclusive results on motivation in XR studies ~\cite{Farr2023,Yang2020}. An extensive post hoc analysis did not reveal any trends, suggesting either no impact on motivation or that the measurement method was not sensitive enough to detect it. Future studies may benefit from alternative approaches to assessing motivation.

\textbf{Perceived Helpfulness:} H6 was not confirmed since there is no significant difference in the VisQ scores between the individual visualisations. The VisQ score of G3 was about 11\% higher (7\% for G2) compared to the control group G1, thus indicating a better-perceived helpfulness when implementing Visual Guidance. The overall trend was a higher perceived helpfulness of the Visual Guidance from G1 to G2 to G3, but the overall differences were insignificant. This was confirmed with the Games-Howell post hoc test. Considering the aforementioned trend and the large standard deviation of G1, adding more participants might lead to significant results, but this remains to be investigated. 

\subsection{Practical Takeaways}
The combination of VG elements with the OAF appeared to effectively reduce TTC by 31\%, but increased the number of mistakes, indicating a trade-off between speed and accuracy. For accuracy-critical tasks, additional safeguards like \textit{poka-yoke}, a strategy helping to prevent mistakes by making them either impossible or at least easy to detect~\cite{robinson1997pokayoke}, should be considered to mitigate errors caused by eliminating the mental rotation process.

Implementing XRVG may compensate for incomplete or confusing instructions. While multiple control participants reported slightly mismatched colors between virtual and physical parts, this challenge was not present for G2 and G3, as Object Identification (OI) VG elements directly pointed out the correct parts. This eliminated the need for users to rely on color comparisons or other selection mechanisms.

Standard AR HMDs remain prone to visual misalignments, tracking errors, and visual artifacts, which users found disruptive~\cite{Hnemann2023}. To improve XRVG resilience, practitioners should enhance tracking accuracy, implement poka-yoke-inspired features tailored to each step~\cite{robinson1997pokayoke}, and provide reliable fallback options, such as alternative interaction methods (e.g., voice commands or "gaze and commit") to address issues like unreliable navigation menus.

The OAF significantly improved TTC, supporting the theoretical model that VG elements primarily aid cognitive processes like part selection while potentially hindering physical execution (see \autoref{OAF}). The findings suggest a good balance between no VG and constant VG for maximizing user performance. Displaying VG elements briefly and then removing them enhances performance, suggesting the need for further research into mitigating occlusion, refining the OAF, and understanding the nuanced relationship between VG elements and user performance.


\subsection{Limitations and Opportunities for Future Research}

Bias stemming from the researcher's presence in the controlled environment to count the mistakes (the ‘Hawthorne Effect') cannot be ruled out~\cite{McCarney2007}. This may have contributed to increased task execution speed and mistakes.

Furthermore, despite the ex-ante power analysis, the achieved statistical power for usability was only 72\%, which did not cross the intended threshold of 80\%. While an ex-post examination on possible type I and II errors did not yield any reason for concern, aiming for a larger sample size in future studies may help to increase the statistical power of the results obtained.

Lastly, the potential misalignment and discolouring of the visual cues, and thus inaccurate or even misleading representation of the situated Visual Guidance represents one of the key limitations of the present study. While the HMD was calibrated and checked for accuracy before each run, anchor shifting or tracking errors may have occurred during the experiment and might not always have been picked up by the research team. Additionally, the finger tracking required to press the ‘next’ button turned out to be highly sensitive to both lighting conditions and skin colour, which resulted in participants having to repeatedly attempt to skip to the next step on several occasions, slightly impacting TTC in the process. The recommendation for future studies is to use an experimental design that requires less accuracy for both tracking and placement than the setup used in the present study.

\section{Conclusion}

This study investigated and quantified the impact of XRVG on task completion time (TTC), errors, cognitive load, usability, motivation, and the perceived helpfulness of Visual Guidance in an AR context, while also introducing and testing a new Occlusion Avoidance Feature (OAF). Results showed a 31\% reduction in TTC but a 380\% increase in errors compared to the control group. Cognitive load and motivation remained consistent across groups, while usability significantly improved with VG. However, this usability boost did not explain the observed variations in TTC or errors.

Contrary to initial hypotheses, cognitive load, usability, and motivation showed no significant impact on observed performance. While usability was notably higher with VG implemented compared to the control group, cognitive load and motivation remained unchanged across groups. Importantly, the increase in usability did not account for the differences in TTC or errors. We further explored the practical application of the XRVG framework in an AR context,  with participant feedback supporting its three dimensions -- Gaze Guidance, Object Identification, and Action Guidance -- and demonstrating the OAF's effectiveness in mitigating occlusion challenges. These findings suggest a complex relationship between Visual Guidance and user performance, as XRVG can enhance cognitive picking and placing processes but may disrupt motoric execution if not optimally implemented.

In conclusion, this study offered new insights into the XRVG framework and OAF's potential to improve performance in assembly tasks, while also revealing an intricate interplay between Visual Guidance, mistakes made, and the underlying human factors. Despite uncertainties about the role of human factors, the strong impact of XRVG on performance remains a key contribution of this study and warrants the need for further research into its underlying mechanisms and optimal implementation for knowledge-intensive procedural tasks.

\begin{acks}
The authors wish to thank K. E. Ruf for her invaluable suggestions and support during this research and the development of this publication.
Additionally, we would like to thank the reviewers for their helpful feedback and the Harvard i-lab for hosting us.
This research is supported in part by the RADMA Association, the Hanns Seidel Foundation, and the Cambridge Trust.
\end{acks}

\bibliographystyle{ACM-Reference-Format}
\bibliography{sample-base}

\end{document}